# Skin dose in breast radiation therapy: Monte Carlo calculations from deformed vector fields (DVF)-driven CT images

N. Arbor[1], L. Bartolucci[2], B. Dai[3], H. Elazhar[2], P. Galmiche[3], D. Jarnet[2], Michel De Mathelin[3], P. Meyer[2], H. Seo[3]

[1]Université de Strasbourg, CNRS, IPHC UMR 7178, F-67000 Strasbourg, France
[2]Département de Radiothérapie, ICANS, Strasbourg, France
[3]Laboratoire ICube, UMR7357, Strasbourg, France

**Abstract** Skin dose in radiotherapy is a key issue for reducing patient side effects, but dose calculations in this high-gradient region remains a challenge. To support radiation therapists and medical physicist in their decisions, a computational tool has been developed to systematically recalculate the skin dose for each treatment session, taking into account changes in breast geometry. This tool is based on Monte Carlo skin dose calculations from deformed vector fields (DVF)-driven CT images, using 3D-scanner data. The application of deformation fields to the initial patient CT, as well as the validation of Monte Carlo skin dose calculations and comparison with TPS algorithm, are presented. The tool is currently being evaluated using data from a clinical study involving 60 patients (MorphoBreast3D).

## 1 Introduction

Breast cancer is the most common cancer among women in France, with around 60,000 new cases and over 10,000 deaths per year. Some 85% of patients are currently treated with radiotherapy [1]. Despite continuous improvements in techniques to optimize dose distribution, i.e. increase the precision of radiation targeting while minimizing exposure to normal tissues, a significant number of patients still suffer from acute toxicity (burns) to the skin and subcutaneous tissues. In order to reduce these side effects, one needs to better estimate the exacts dose delivered to the skin during irradiations. However, calculating the dose distribution in this very high-gradient region is complicated by the combination of the lack of precision of treatment planning systems (TPS) at the air-tissue interface [2], the compromise to be made on the calculation grid size in order to remain within acceptable calculation times, and the inevitable changes in breast geometry during and throughout treatment (breathing, breast swelling between fractions) [3].

Radiotherapy centers have developed protocols to limit the possible side-effects of inaccurate surface dose estimation, such as modifying the patient's target or body volume margins [4], or using a bolus during the optimization process to spare dose build-up to the skin. But the main difficulty encountered by medical physicists is the lack of tools available for determining the optimal solution for each patient, especially when considering intensity modulated techniques with regard to changes in breast geometry during treatment.

In this context, the Strasbourg Europe Cancer Institute (ICANS), the ICube laboratory, and the Hubert Curien Interdisciplinary Institute (IPHC) are collaborating on the development of a computational framework dedicated to skin dose calculation in breast radiotherapy. This framework combines a Monte Carlo dose calculation tool and a breast deformation algorithm by using 3D surface data recorded during treatment sessions.

This framework allows:
1) to evaluate the necessity for recalculating the treatment plan, based on the production of deformed CT images that reproduce the inter-fraction evolution of breast geometry;
2) to exploit the precision of the Monte Carlo calculation to estimate the dose delivered in the 1st millimeter of skin. This complements and evaluates the calculations of the TPS algorithms, thereby improving the understanding of the induction of side effects introduced by the breast deformation.

After introducing the general framework architecture, we present some measurements and Monte Carlo calculation of skin dose, the method to compute CT images of deformed breast by using 3D surface measurements, as well as the results we obtained on the first patients as part of a clinical study.

## 2 Materials and Methods

The skin is a complex organ, characterized by a layered structure with different cell types. The thickness of the epidermis varies between 0.05 and 1.5 mm depending on the anatomical region. The International Commission on Radiological Protection (ICRP) and the International Commission on Radiological Units and Measurements (ICRU) recommend a thickness of 0.07 mm for the practical study of the dose to the skin.

70 μm remains a precision that is difficult to access with current clinical dose calculation algorithms, thus motivating the use of Monte Carlo calculation to address the problem of skin dose in breast radiotherapy. Our dose calculation tool is based on the GATE Monte Carlo software [5], and has been previously validated on a 6 MV TrueBeam STX accelerator [6]. This tool uses the phase spaces provided by the manufacturer and the information available in the DICOM-RT files (image, plan, structure, dose) to calculate the dose delivered to the patient. A digital phantom of the patient is created from the scanner images and the HU-materials conversion method proposed by Schneider et al. [7].

The dose to the skin is due to low energy photons scattered from the accelerator head, photons backscattered into the medium, and contamination electrons produced in the accelerator head and the head-patient interface (air, accessories). Significant validation work was therefore carried out to ensure the good accuracy of the Monte Carlo dose calculation in the first millimeters of tissue. Contamination electrons, which can contribute up to nearly 50% of the dose delivered to the skin depending on the configurations, have been particularly studied from Varian phase spaces.

A clinical study carried out at ICANS in 2019-2020 enabled the acquisition of 60 patient datasets (*MorphoBreast3D*). These data contain CT images of the patients used for planning (with contouring of the organs including the skin), to dose maps calculated by the TPS, and a set of 3D surfaces meshes of the patient bust acquired at different sessions of treatment using a Artec Eva 3D hand-held scanner (Figure 1).

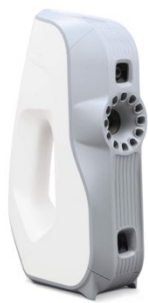

*Figure 1. Artec Eva 3D scanner*

The objective is to use the 3D surfaces recorded during a session in order to deform the initial CT image of the patient and to obtain a new CT image representing the exact breast geometry during the session. A per-vertex correspondence has been found between a 3D surface pair, one acquired from the CT scan ($M_{init}$), and another during a session ($M_s$). Then a deformation vector field is calculated by aligning the two surfaces and connecting the vertices in correspondence. The displacement vector field is then transferred to the CT image by rigidly aligning the surface $M_{init}$ and the CT image. We finally obtain the deformed CT image via the use of radial basis functions (RBF), whose point specifications are defined by the displacement vector. The latter step uses voxelization and interpolation techniques to ensure the overall consistency of the distorted CT image which will serve as input for dose calculations.

CT images containing deformed breast geometry serve as input to both TPS global dose calculations and GATE Monte Carlo skin dose calculations. The use of TPS to recalculate the dose map on the exact geometry of the breast during the session should enable medical physicists to estimate whether a replanning of the treatment is necessary, by examining deviations from the initial dose plan. The calculation of dose to the skin via our Monte Carlo tool is used to provide additional information to the TPS calculations for the first millimeter of tissue, and to precisely study the impact of deformations on the distribution of dose to the skin, such as under or overestimation, hot spots.

## 3 Results

Figure 2 shows an example of a deformation vector field calculated on a patient's 3D surface mesh $M_{init}$.

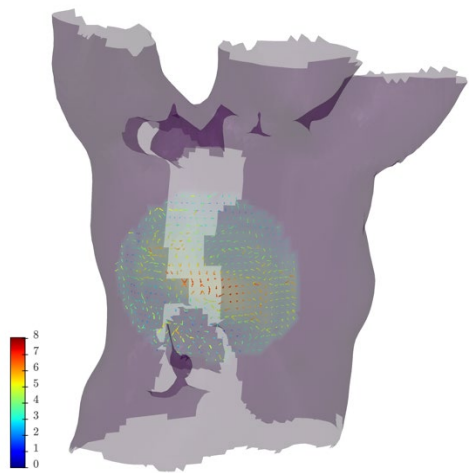

*Figure 2. Deformation vector field*

From this deformation field, our algorithm produces a CT image representing the new geometry of the deformed breast for a given treatment session (Figure 3).

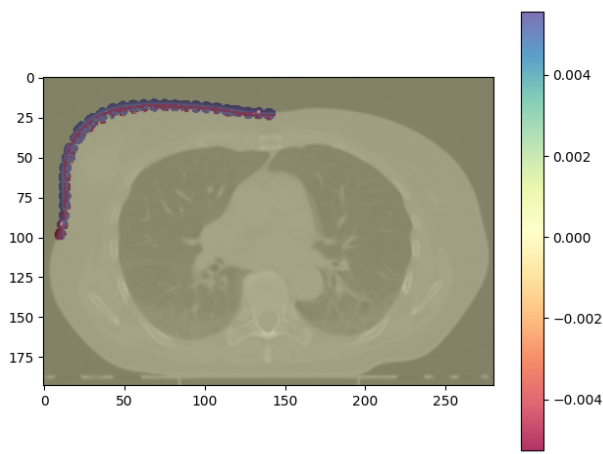

*Figure 3. Breast deformation (colored voxels (a.u.)) from pre-treatment patient CT*

Before applying Monte Carlo calculations to initial and deformed CT images, experimental validations of skin dose have been carried out on PMMA phantoms for various treatment plans. Monte Carlo dose calculations in the first millimeter have been compared to EBT3 (Ashland, Bridgewater, NJ, USA) Gafchromic films [8], demonstrating the good calculation performance of GATE Monte Carlo algorithm at this millimeter scale (Figure 4).

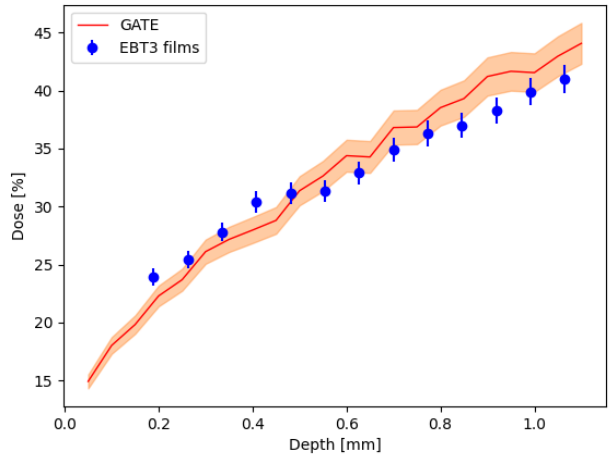

*Figure 4. Depth dose comparison between GATE (Monte Carlo) and EBT3 film measurements (PMMA phantom)*

We have currently tested our method on two patients of the clinical trial. For these patients, a mean breast deformation field length of about 5 mm has been obtained. For each patient, dose maps were calculated by using the Acuros TPS and GATE MC tool for five deformed CT images (D+5, D+10, D+15, D+20, and D+25), and subsequently compared to the dose map from the original treatment plan.

## 4 Discussion

These first tests of a skin dose calculation framework, carried out as part of this work, are very encouraging. The study will continue its systematic application to all 60 patients who participated in the clinical study. The ultimate goal is then to exploit the developed framework in several complementary applications:
a) verification of the dose calculation after each session, taking into account the exact geometry of the breast, offering both to the radiation therapists and medical physicists a decision support tool to determine the necessity of treatment replanning;
b) set up research work on the precision of surface dose calculations from TPS algorithms (contour, thresholding) by a systematic comparison with MC calculations
c) provide data for a new clinical study on understanding the side effects of breast radiotherapy depending on the type of in-therapy deformation (shrink, inflation, folds)

Other tools exist today that allow recalculation of the treatment plan based on breast deformations at each session. These tools are mainly based on the use of CBCT images taken at the beginning of the session for the repositioning of the patient. One of the specificities of the proposed framework compared to these tools consists of using 3D surfaces obtained from an optical scanner. Its application does not require an additional X-ray image, relying solely on the initial CT, thus avoiding extra imaging doses for the patient and adhering to the ALARA principle for patient radiation protection. Another advantage of the algorithm developed in this work is its applicability to exploit images provided by the SGRT (Surface Guided Radiation Therapy) systems, which are increasingly present in treatment centers.

## 5 Conclusion

Skin dose calculation in radiotherapy is a key issue for reducing patient side effects. A framework has been developed to enable systematic recalculation of the skin dose at each session, taking into account breast deformations from 3D-scanner images. This framework enables Monte Carlo skin dose calculations based on deformed vector fields (DVF)-driven CT images. The first patient results, as well as the availability of SGRT images, point to the future clinical use of this non-ionising method to support radiation therapists and medical physicists in their replanning decisions during treatment, and to produce useful data for improving treatment precision and better understanding the induction of side effects.